\documentclass[12pt]{article}

\usepackage{amssymb}
\usepackage{amsfonts}
\usepackage{amsmath}

\usepackage{graphics}
\usepackage{graphicx}
\usepackage{float}     

\usepackage{titletoc}
\usepackage[raggedright,pagestyles]{titlesec}

\usepackage{hyperref}

\usepackage{caption2}

\usepackage{geometry}

\usepackage{lastpage}

\titleformat{\section}{\raggedright\normalsize\scshape}{\bf\normalsize{\S\,\thesection}.}{1em}{}

\begin{document}


\begin{center}
\large{Conjecture on imminent earthquake prediction\\
 --- from shaving foam to cloud patterns}

\vfill

\begin{small}
Xin LIU

\it{School of Mathematics and Statistics, University of Sydney,
NSW 2006, Australia\\
Email: topkidlx@gmail.com}
\end{small}

\end{center}

\vfill

\noindent \rule{\textwidth}{0.1pt}
\begin{center}
\sc Abstract
\end{center}
\noindent A conjecture on imminent earthquake prediction is
presented. Drastic geological deformations of crustal rock strata
taking place immediately (hours/days) before an earthquake may
cause fast air or gas emission/absorption vertically in between
ground and sky. I conjecture, inspired by an observation of
strange patterns appearing on shaving foam, that this fast
movement of air fluid may produce unusual cloud patterns at
interfaces between atmosphere levels. This air movement is
vertical and drastic, different from the horizontal and moderate
meteorological air movement, hence its caused cloud patterns are
expected to be different from meteorological cloud patterns. This
provides a possible origin for the so-called \textit{earthquake
cloud}. Recognition of different earthquake cloud patterns may
provide a practical way to estimate location, magnitude and
strength of geological deformations of rock strata, and hence a
method with support of physics for imminent earthquake prediction.
In the end of this paper an experiment has been designed to test
the conjecture.

\noindent \rule{\textwidth}{0.1pt}

\vfill


\section{Introduction} \label{Sect01}

Imminent earthquake prediction is a difficult problem. Only
long-term (years to decades) and medium-term (months to years)
predictions are regarded possible at present, while imminent/short
term predictions are commonly thought to be negative
\footnote{Approaches seismologists have used to investigate
earthquakes include the researches on seismicity patterns, crustal
movements, ground water level in wells, radon or hydrogen gas
emissions, changes of seismic wave velocities, electromagnetic
fields (seismo-electromagnetics), large-scale changes in soil
temperature, changes in ion concentration in the ionosphere, and
so on.}. See
\href{http://en.wikipedia.org/wiki/Earthquake_prediction}{\it{Wikipedia:
earthquake prediction}}, and
Refs.\cite{MainNature1997,MainNature1999,GellerScienc1997,BBCnews2008,EqReview1,EqReview2}.
The conjecture and research proposal of this paper are a
suggestion and attempt only.

This paper is arranged as follows. The conjecture will be outlined
in \S\ref{Sect02}, of which the central idea is \textit{vertical
fast air emission/absorption causing cloud patterns}. This idea is
inspired by an observation of strange patterns on shaving foam of
\S\ref{Sect03}. Discussions on geological deformations of crustal
rock strata and the induced vertical fast air movement will be
presented in \S\ref{Sect04}. This air movement may lead to
formation of \textit{earthquake cloud} patterns, hence a
literature survey for earthquake cloud, including some sorts of
explanations, is presented in \S\ref{Sect05}. Potential evidences
for the conjecture, as well as the induced side-effects, will be
listed in \S\ref{Sect06}. An experiment designed to test the
conjecture will be given in \S\ref{Sect07}. Finally, the paper
will be summarized in \S\ref{Sect10}, followed by some remarks in
\S\ref{Sect11}.


\section{Conjecture and research proposal} \label{Sect02}

A conjecture is made on imminent/short term earthquake prediction
based on cloud patterns caused by fast air emission/absorption
between ground and sky:

\begin{enumerate}
\itemsep=2pt
\parsep=1pt
\parskip=0pt

\item Immense volume of air (or gas) stuffs gaps and cracks among
crustal rock strata. Geological deformations could either squeeze
this air out, or produce more room in the rock strata gaps to
absorb in the air from outside.

\item It is a reasonable hypothesis that earthquakes (or, a part
of earthquakes) are three-stage events, from energy preparation to
release \cite{BreakingVideos}:
\begin{enumerate}
\itemsep=2pt
\parsep=1pt
\parskip=0pt

\item \textsf{Long-term preparation (years prior to an
earthquake)}: Gradual geologic motions of tectonic plates push
rock strata to undergo geologic deformations, such as bending,
compression, etc.. Seismological energy is accumulated in this
stage. Geologic deformations become more and more severe when
close to the earthquake, accompanied by occurrence of more and
more minor breakings.

\item \textsf{\textbf{Eve of event} (hours to days prior to the
earthquake)}: This short stage is the final preparation period.
Occurrence of one or two medium-size breakings make the rock
strata suffer \textit{a series of drastic geologic deformations}
within a short time, which will trigger the eventual major
breaking and collapses.

\item \textsf{Occurrence of event (disastrous moment)}: Major
breaking and collapses take place, seismological energy being
released.
\end{enumerate}

\item The \textsf{eve of event} stage is crucial for imminent
earthquake prediction. I conjecture that those \textit{drastic
deformations of rock strata} will cause a huge pressure difference
of air between the two sides of the Earth soil surface, because
the thick soil surface covers the rock strata and acts as an
obstruction of air release (see Fig.\ref{Fig-05} below).
Consequently, vertical high-speed air emission and absorption can
be produced between the ground and sky. In regard to the
observation of the following \S\ref{Sect03}, I conjecture that
this air movement will lead to formation of unusual cloud patterns
on interfaces between atmosphere levels. This provides a possible
origin for the folk-called \textit{earthquake cloud}.

\item This air emission/absorption is vertical and drastic,
different from the horizontal and moderate meteorological
movements of atmosphere, hence its induced cloud patterns are
expected to be different from meteorological cloud
patterns\footnote{Vertical air movements may also appear in
extreme weather situations. However, since earthquakes and extreme
weather are both small-probability events, the probability of
co-existence of them is even smaller.}. Therefore, earthquake
cloud appearing hours/days before an earthquake could be a
candidate for imminent/short term earthquake prediction, with
support of physics.

\item Furthermore, it is thought that different magnitude,
strength and velocity of air emission/absorption could produce
different cloud patterns. Hence the patterns provide a way to
reveal more quantitative information of location, magnitude and
strength of the geologic deformation of rock strata, and hence of
the impending earthquake.

\end{enumerate}

\noindent The conjecture leads to the following \textbf{research
proposal}, which is a study of inverse problem:
\begin{itemize}
\itemsep=2pt
\parsep=1pt
\parskip=0pt

\item[$\triangleright$] \textit{Step 1 (Cloud patterns):}
Distinguishing the cloud patterns of earthquakes from the cloud
patterns of meteorological movements of atmosphere, and further
recognizing different sorts of patterns of earthquake cloud.

\item[$\triangleright$]\textit{Step 2 (Air movement):} Estimating
magnitude, strength and velocity of vertical air emission and
absorption --- establishing workable mathematical/mechanical
models.

\item[$\triangleright$]\textit{Step 3 (Geologic deformations):}
Estimating locations, magnitude and strength of geologic
deformations of rock strata, with the aid of long- and medium-term
predictions of seismology.

\item[$\triangleright$]\textit{Step 4 (Final purpose):} Imminently
predicting earthquakes.
\end{itemize}

\section{Patterns appearing on shaving foam} \label{Sect03}

\begin{figure}[h] \centering
\includegraphics[width=0.55\textwidth]{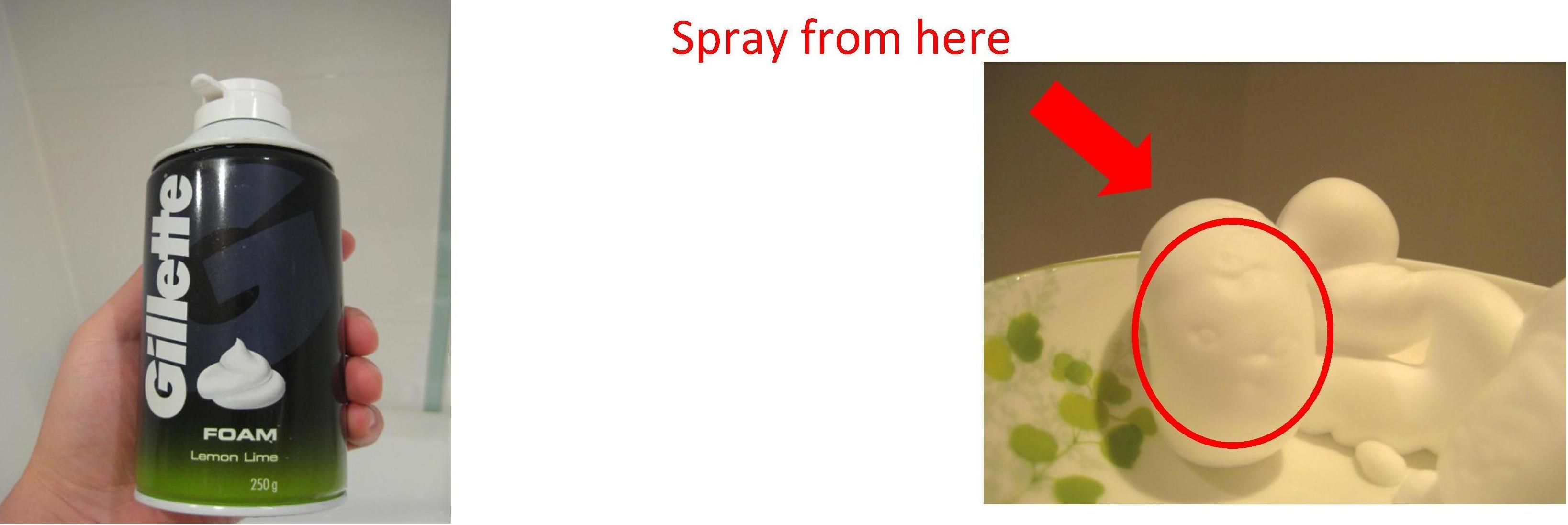}
\caption{Shaving foam sprayed out from an aerosol can.
\textit{Left}: The sample used was a half-filled can of
\textit{Gillette Lemon Lime}, which was shaken 4 or 5 times before
spraying. \textit{Right}: The foam was sprayed out from the can
and patterns appeared on the foam surface.} \label{Fig-01}
\end{figure}

The idea of \textit{air-emission causing patterns} is inspired by
the following observation on strange patterns appearing on shaving
foam when the foam is sprayed out from an aerosol can. See
Fig.\ref{Fig-01}.

Before that observation I took for granted that the sprayed foam
should always be smooth, as in Fig.2; however, it was not the
case. Instead, in some situations, such as ``\textit{half-filled
can plus shaking the can 4 or 5 times}'', patterns appeared on the
sprayed foam. See Figs.\ref{Fig-01} and \ref{Fig-03}.
\vfill

\vfill\begin{figure}[h] \centering
\includegraphics[width=0.17\textwidth]{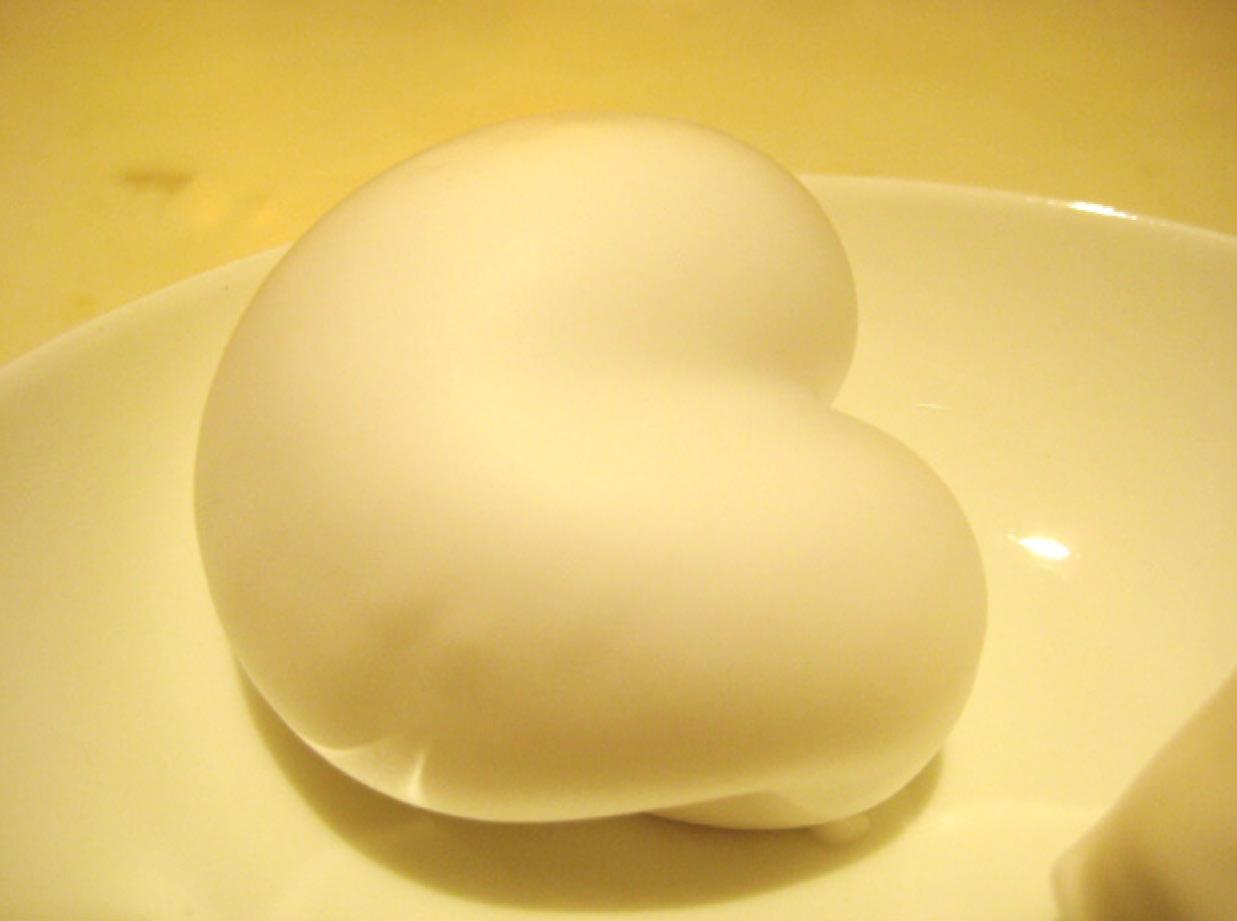}
\caption{Smooth foam containing no patterns.} \label{Fig-02}
\end{figure}

\vfill\begin{figure}[h] \centering
\includegraphics[width=0.9\textwidth]{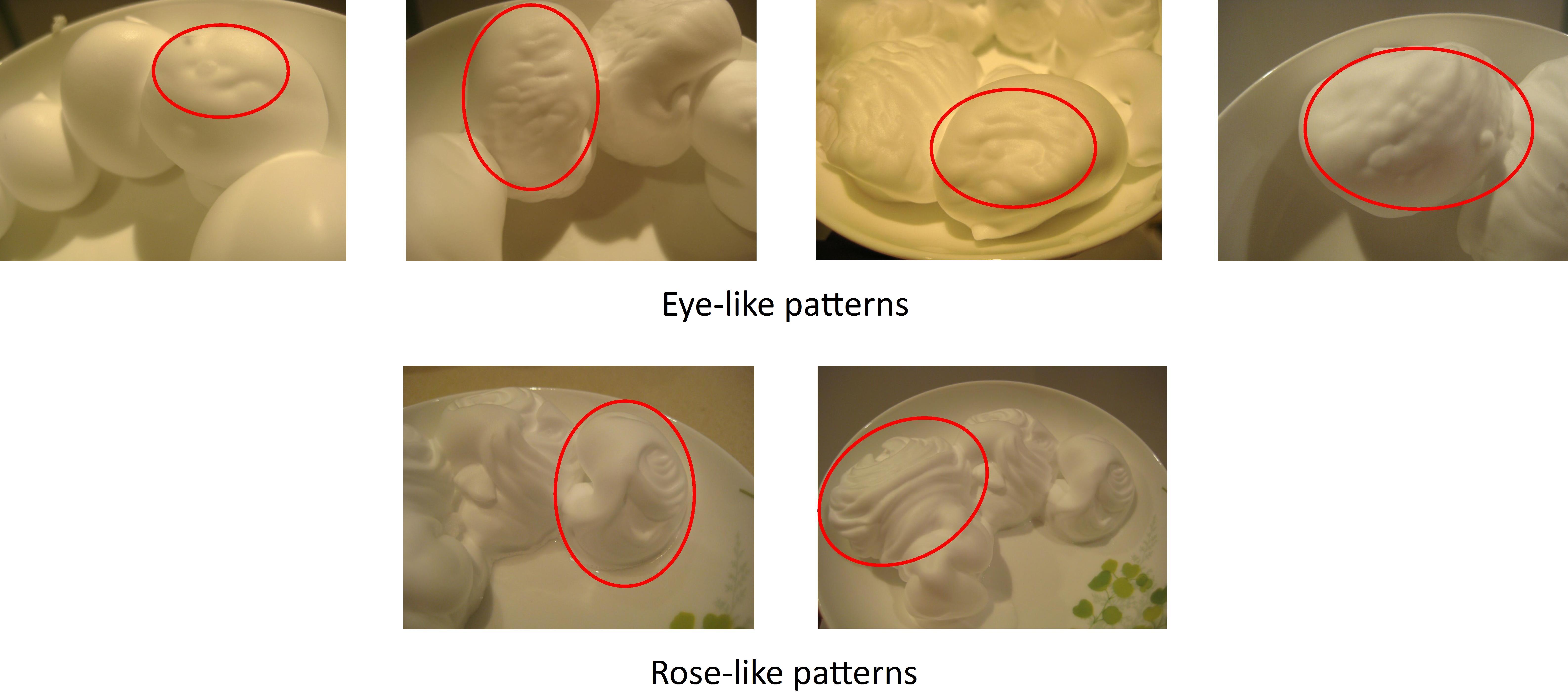}
\caption{More patterns on shaving foam.} \label{Fig-03}
\end{figure}

\noindent In Figs.\ref{Fig-01} and \ref{Fig-03} different patterns
are shown, including the eye-like and the rose-like. I attribute
the formation of these patterns to the fast aerosol spray. This
spray was caused by the pressure difference between the inside and
outside of the aerosol can, while the pattern formation was
governed by some fluid dynamical mechanism that is still unknown
\footnote{For instance, a possible mechanism is that the fast
spray causes oscillation and waves in the aerosol which form the
patterns at the interface between the aerosol and the outer air,
just like sound waves in air which are able to cause ripples when
travelling over smooth water surface.\label{FTNote}}. See
Fig.\ref{Fig-04} below.

\begin{figure}[h] \centering
\includegraphics[width=0.85\textwidth]{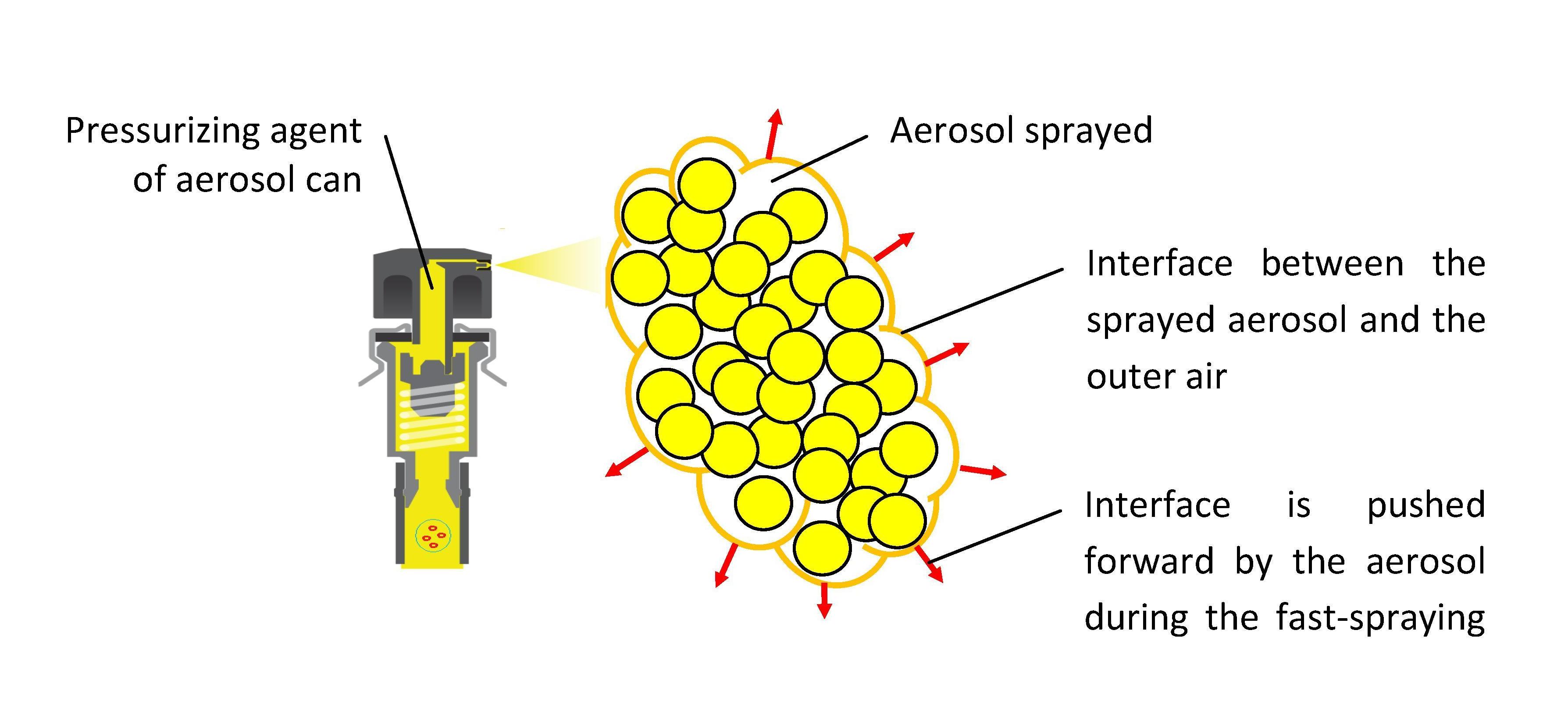}
\caption{Aerosol is fast sprayed out from a can. During the spray
process, the interface between the aerosol and the outer air is
strongly pushed forward by the aerosol, where some unknown fluid
dynamical mechanism causes appearance of strange patterns on the
interface.} \label{Fig-04}
\end{figure}

\vfill

\section{Fast air emission/absorption due to geological deformations} \label{Sect04}

The phenomena observed above could occur on cloud prior to an
earthquake (see Fig.\ref{Fig-05} below). As mentioned in
\S\ref{Sect02}, many an earthquake has a crucial short stage
immediately before the major breaking moment, where drastic
geological deformations take place. The thick soil surface
covering the rock strata causes a huge pressure difference of air
between the two sides of the soil surface, which leads to vertical
fast air emission and absorption. The bigger the deformations are,
the more violent the air emission and absorption are. \vfill
\newpage

\begin{figure}[H]
\includegraphics[width=1.03\textwidth,height=0.90\textheight]{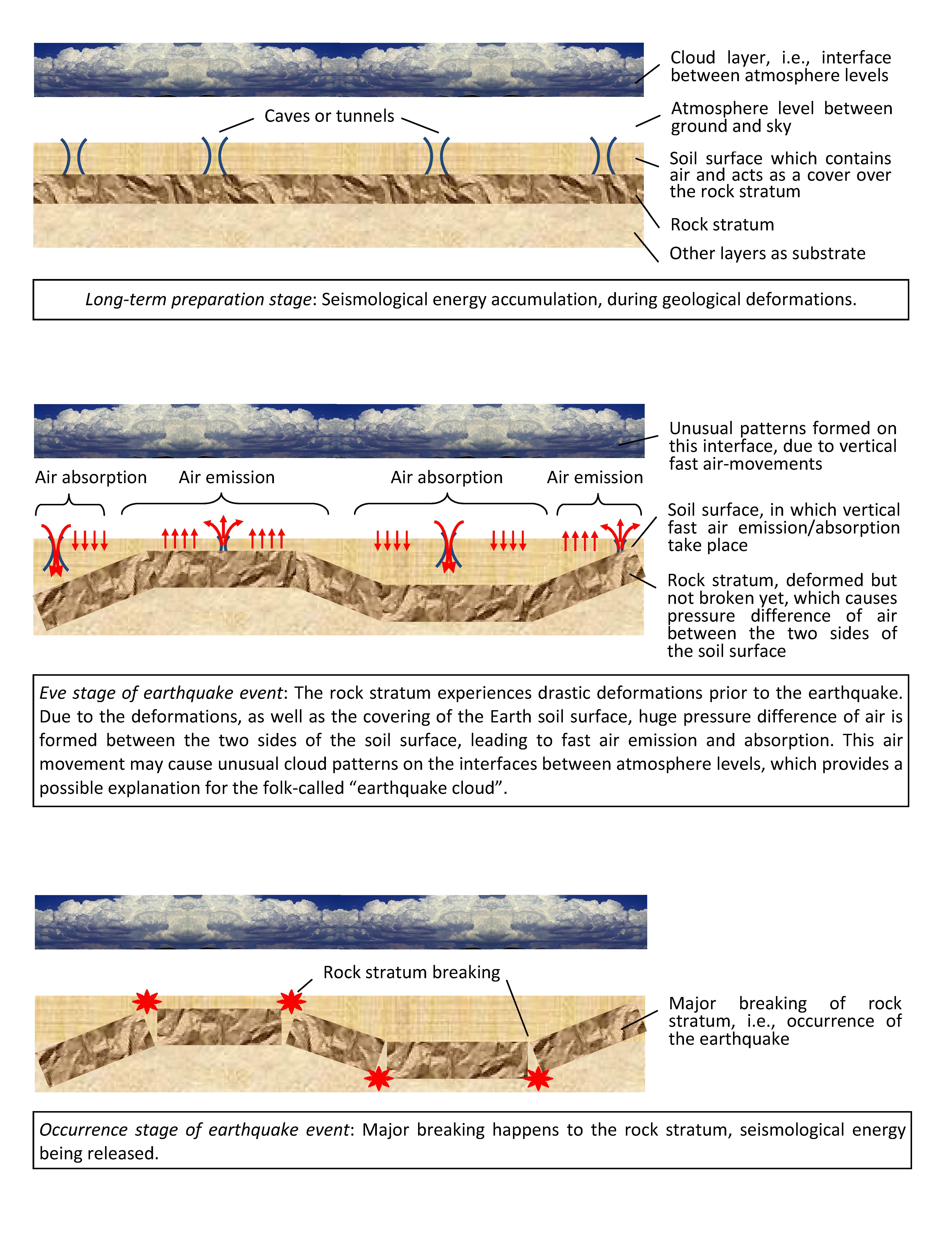}
\caption{Three stages of earthquake event. The second stage,
\textit{eve of event}, is crucial for imminent earthquake
prediction, where vertical fast air emission and absorption could
form unusual cloud patterns on interfaces between atmosphere
levels.} \label{Fig-05}
\end{figure}

\section{Literature survey of earthquake cloud} \label{Sect05}
\vspace*{-8mm}
\begin{figure}[H] \centering
\includegraphics[width=1.0\textwidth]{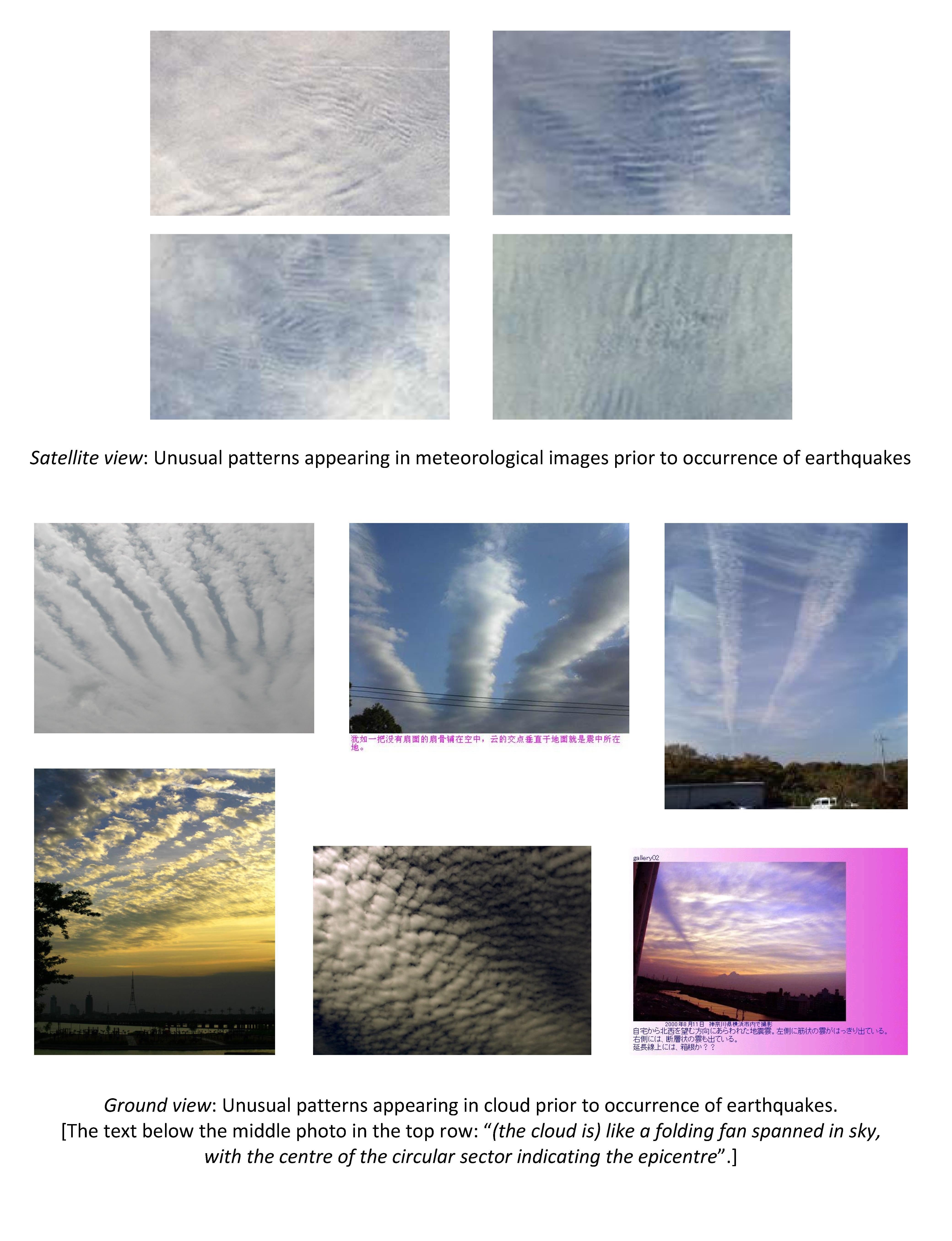}
\caption{Photos of the so-called earthquake clouds from different
angles of view. (Meteorological satellite images are from
Ref.\cite{TerraResLAPark}.) } \label{Fig-06}
\end{figure}

Earthquake cloud has not been commonly accepted by the scientific
community as a sign of impending earthquakes \cite{WikiEqCloud}.
There are diametrically opposite opinions: some people treat
simultaneous occurrence of unusual-looking cloud and an earthquake
as a coincidence; but some people believe the strange-looking
clouds in Fig.\ref{Fig-06} are associated with seismic events.

Earthquake cloud has been observed and studied for years, based on
which some earthquake predictions have been made. For instance, a
Chinese folk-scientist named Z. Shou claimed that he had made
dozens of earthquake predictions based on cloud patterns in
satellite images, with a 68\% accuracy
\cite{ShouZ2005,ShouZwebsite}. He identified five types of
earthquake cloud --- line-shaped, feather-shaped, lantern-shaped
clouds, etc. --- and claimed that appearance of any one of these
clouds indicates an impending earthquake to occur within several
hours to 103 days (averagely 30 days).

Some explanations for earthquake cloud are the following
\cite{WikiEqCloud}:
\begin{itemize}
\item[$\triangleright$] \textit{Heat-flow paradox}

\textit{Claim}: Deformations and motions of rock strata cause rock
frictions, and hence produce a vast amount of thermal energy. This
energy heats water to 1500$^{\circ}$C such that hot vapor obtains
an updraft to form earthquake cloud. This explanation is held by
Z. Shou.

\textit{Deficiency}: If this is true, such hot vapor should have
been noticed by human beings, but that is not the case. In
addition, geophysicists have conducted experiments and pointed out
that this heating is only 4$^{\circ}$C or so, not enough to
produce earthquake cloud \cite{TerraResLAPark}.

\item[$\triangleright$] \textit{Effect of piezoelectricity}

\textit{Claim}: Piezoelectricity occurring inside the Earth causes
local variation of geomagnetic field, which leads to variation in
electromagnetic fields in sky and thus forms earthquake cloud.

\textit{Deficiency}: This explanation is unconvincing.
Piezoelectricity is a kind of electromagnetic effect, which is far
too weak to affect motions of atmosphere. Moreover, if the
piezoelectricity is strong enough to drive motions of air, then
horrible electromagnetic noise must have been created to disrupt
communications and destruct electronic devices. But that is not
the case apparently.
\end{itemize}

\noindent From my point of view, only the explanations mostly
based on mechanical reasons are acceptable. My explanation is the
following, associated with the conjecture of \S\ref{Sect02}:
\begin{quote}
An earthquake event has a final preparation stage immediately
before the major breaking moment, where the drastic geological
deformations cause vertical fast air movements which produce
earthquake cloud. Some unknown mechanism of fluid dynamics takes
effect in this process, similar to that of the footnote on Page
\pageref{FTNote}. I.e., fast air movements produce oscillation and
(sound) waves in the atmosphere when the waves propagate to the
interfaces between atmosphere levels. The sound waves produced by
the air movement could be ultra-, acoustic or infrasound.
\end{quote}

\vfill

\section{Potential evidences and side effects} \label{Sect06}

Vertical fast air emission/absorption via the Earth soil surface,
either by rushing through caves and tunnels or by penetrating
through soil (as shown in Fig.\ref{Fig-05}), could have potential
evidences and induced side effects:

\begin{itemize}
\item[$\triangleright$] \textit{Gas emission prior to volcanic
eruption}

It is instructive to refer to the phenomenon of gas emission
before volcanic eruption, which is an example of vertical fast air
emission associated with geological motions of rock strata of
tectonic plates.

\item[$\triangleright$] \textit{Sound effect}

Air emitted through caves, tunnels or soil may produce sound waves
of high, intermediate or low frequency, as a horn does when air
passing through it. Hence ultra-, acoustic or infrasound could be
heard by human ears or devices before earthquake events.
\footnote{This sound effect might be different from the so-called
phenomenon of \textit{earthquake sound} which takes place seconds
or minutes ahead of earthquake occurrence.}

\item[$\triangleright$] \textit{Light effect}

High-speed air movement may cause frictions among ionized air
masses which result in the phenomenon of lightening. This could be
an origin for the so-called \textit{earthquake light}.

\item[$\triangleright$] \textit{Atmospheric ionization}

According to Wadatsumi (Okayama University of Science, Japan)
\cite{Wadatsumi}, ionization of atmospheric aerosol rises
remarkably prior to an earthquake, which results in a phenomenon
that the color of sky turns red. Human beings may feel
uncomfortable and depressed in such a gas emitted from the inside
Earth.

\item[$\triangleright$] \textit{Release of radon gas}

Radon (\textsf{Rn}, atomic number 86) is a radioactive, colorless,
odorless, tasteless noble gas, which is thought to be released
from fault zones prior to earth slipping. Researchers have
investigated changes in groundwater radon concentrations for
earthquake prediction.

See the work of the research group led by G. Charpak (Nobel
laureate in physics, 1992) in \cite{PhysicsWorldNews}.

\item[$\triangleright$] \textit{Anomalous animal behaviors}

Animals, especially those burrowing and underground-living
animals, will probably be disturbed by the bad smell or even toxic
gas emitted from the inside Earth.

\end{itemize}

\vfill
\section{Design of experiment to test vertical fast air movement} \label{Sect07}

An experiment is designed to test the conjecture of \S\ref{Sect02}
and \S\ref{Sect04}:
\begin{figure}[H] \centering
\includegraphics[width=0.8\textwidth]{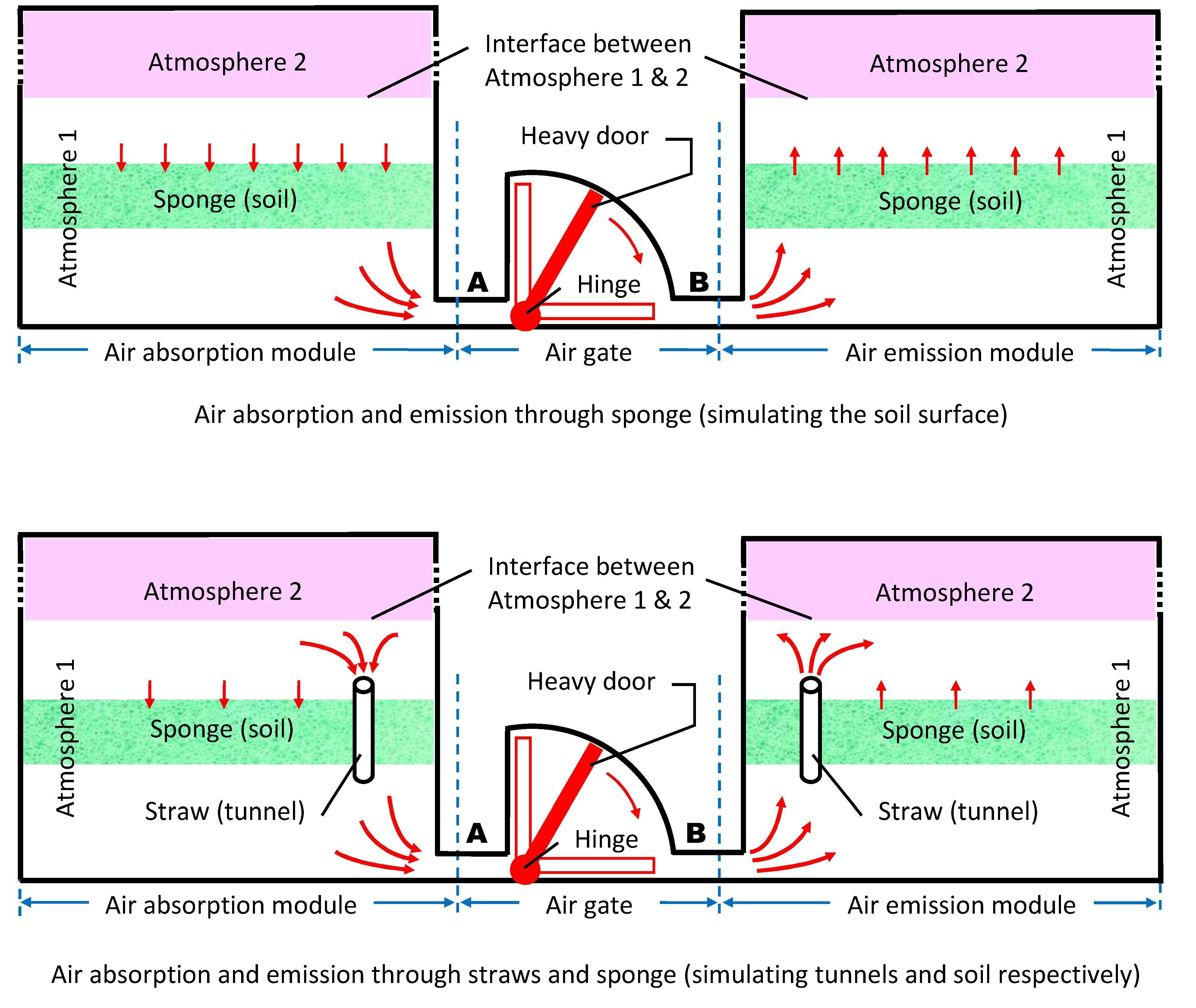}
\caption{Experiment designed to observe patterns on atmosphere
interfaces formed by air emission/absorption. In this experiment
sponge is used to simulate the Earth soil surface, and straws to
simulate caves and tunnels.} \label{Fig-07}
\end{figure}

\begin{itemize}
\itemsep=5pt
\parsep=2pt
\parskip=2pt

\item[$\triangleright$]\textit{Upper part of Fig.\ref{Fig-07}
--- air penetrating through soil}

This set-up is divided into three modules:
\begin{itemize}
\itemsep=0pt
\parsep=5pt
\parskip=2pt

\item \textit{Middle --- Air-gate module}: It is an empty
quarter-cylinder, which has two entries marked as Gate \textbf{A}
and \textbf{B}. This module contains an air marked as Atmosphere
1. A standing heavy door is placed in this module, which can fall
down by rotating about a hinge. This door serves as an air piston.

\item \textit{Left --- Air absorption module}: It is connected to
the Air-gate via Gate \textbf{A}. This module contains not only
Atmosphere 1, but also another air marked as Atmosphere 2 located
on top of Atmosphere 1. Atmosphere 2 is lighter in density. These
two Atmospheres should be carefully chosen, such that they do not
mix up with each other and have a clear interface on which the
desired patterns can be demonstrated. In the middle of this module
a layer of sponge is placed to play the role of the Earth soil
surface.

\item \textit{Right --- Air emission module}: It is connected to
the Air-gate via Gate \textbf{B}. This module also contains
Atmosphere 1 and 2, with the latter on top of the former, to form
a clear interface. In the middle of this module there is also a
sponge.
\end{itemize}

In the Middle module, when the piston door falls down, air is fast
absorbed in from the Left module via Gate A. At the same time, in
the Left module, the part of Atmosphere 1 above the sponge is
absorbed down to penetrate through the sponge. During this
process, \textit{patterns are expected to appear on the interface
between Atmospheres 1 and 2 in the Left module}.

Similarly, when the piston door of the Middle module falls down,
air is fast pumped out from the Middle to the Right module via
Gate B. At the same time, in the Right module, the part of
Atmosphere 1 beneath the sponge is emitted up to penetrate through
the sponge, and \textit{patterns are expected to appear on the
interface between Atmospheres 1 and 2 in the Right module}.

The purpose of this experiment is to observe the patterns formed
on the interfaces, with the presence of the obstructive sponge
which simulates the Earth soil surface.

\item[$\triangleright$]\textit{Lower part of Fig.\ref{Fig-07}
--- air spraying through caves and tunnels or penetrating through soil}

This set-up is almost the same as the upper part. The only
difference lies in that several straws are placed in the sponge,
for the purpose of simulating caves and tunnels of the soil
surface. They provide another air passageway in addition to the
sponge, such that air can either rush through the straws or
penetrate through the sponge. It is expected that different
patterns could be obtained on the interfaces between Atmospheres 1
and 2 in the Left and Right modules.

\end{itemize}

\noindent In this experiment we can choose different-sized
Air-gate modules, to produce air movement with different
magnitude, strength and velocity. We expect to obtain different
patterns on the interfaces. Moreover, different choices of
Atmospheres 1 and 2 are expected to bring extra alteration to the
patterns obtained.

\vfill

\section{Summary} \label{Sect10}

In this paper a conjecture is made on imminent earthquake
prediction based on cloud patterns. In \S\ref{Sect02} the contents
of the conjecture are outlined. In \S\ref{Sect03} an observation
of the strange patterns appearing on shaving foam are presented.
In \S\ref{Sect04} it is illustrated that drastic geological
deformations of rock strata, taking place immediately (hours/days)
before an earthquake, may cause fast air emission/absorption
through the Earth soil surface, vertically in between ground and
sky. Inspired by the observation of \S\ref{Sect03}, it is
conjectured that this fast movement of air fluid may produce
unusual cloud patterns at interfaces between atmosphere levels.
Different from the horizontal and moderate meteorological air
movement, this air movement is vertical and drastic, hence its
caused cloud patterns are expected to be different from
meteorological cloud patterns. This provides a possible origin for
the so-called \textit{earthquake cloud}. Recognition of different
earthquake patterns could provide a practical way to estimate
magnitude, strength and location of geological deformations of
rock strata, and hence a physical method for imminent earthquake
prediction. In \S\ref{Sect05} literature survey and explanations
for earthquake cloud are presented. In \S\ref{Sect06} potential
evidences of the vertical fast air emission/absorption and some
induced side effects are listed. Finally in \S\ref{Sect07} an
experiment is designed to test the conjecture.

\vfill

\section{Remarks} \label{Sect11}

\begin{itemize}

\item[$\triangleright$] In \S\ref{Sect03} an observation of the
strange patterns on shaving foam has been shown. The sample used
was \textit{Gillette Lemon Lime}, and the experimental condition
is ``\textit{half-filled can plus 4--5 shakes}''. Another sample,
\textit{Gillette Sensitive Skin} foam, has been tried as well, for
which it is found that the condition of producing patterns is
different, and the patterns obtained are not so clear. This
implies that the ingredients of aerosol may somehow affect the
patterns formed.

\item[$\triangleright$] The personal research areas of the author
are \textit{topological fluid mechanics} and \textit{topological
quantum field theory}, far from the topic of this paper. This
paper stems from my personal interest; it is an attempt to share
ideas with colleagues. There is not much mathematics in this
paper, and the ideas could be incorrect.

\end{itemize}

\vfill
\begin{small}

\end{small}

\vfill

\end{document}